\documentclass[pre,twocolumn,showpacs,amsmath,amssymb,floatfix]{revtex4}

\usepackage{graphicx}
\usepackage{dcolumn} 
\usepackage{bm}      
\usepackage{color}

\begin{document}

\title{Non-local interactions in hydrodynamic turbulence at high Reynolds
       numbers: \\ the slow emergence of scaling laws}
\author{P.D. Mininni$^{1,2}$, A. Alexakis$^3$, and A. Pouquet$^2$}
\affiliation{$^1$ Departamento de F\'\i sica, Facultad de Ciencias Exactas y
         Naturales, Universidad de Buenos Aires, Ciudad Universitaria, 1428
         Buenos Aires, Argentina. \\
             $^2$ NCAR, P.O. Box 3000, Boulder, Colorado 80307-3000, U.S.A.\\
             $^3$ Laboratoire Cassiop\'ee, Observatoire de la C\^ote d'Azur,
         BP 4229, Nice Cedex 04, France.}
\date{\today}

\begin{abstract}
We analyze the data stemming from a forced incompressible
hydrodynamic simulation on a grid of $2048^3$ regularly spaced
points, with a Taylor Reynolds number of $R_{\lambda}\sim 1300$. The
forcing is given by the Taylor-Green flow, which shares similarities
with the flow in several laboratory experiments, and the computation
is run for ten turnover times in the turbulent steady state. At this
Reynolds number the anisotropic large scale flow pattern, the
inertial range, the bottleneck, and the dissipative range are
clearly visible, thus providing a good test case for the study of
turbulence as it appears in nature. Triadic interactions, the
locality of energy fluxes, and structure functions of the velocity
increments are computed. A comparison with runs at lower Reynolds
numbers is performed, and shows the emergence of scaling laws for
the relative amplitude of local and non-local interactions in
spectral space. The scalings of the Kolmogorov constant, and of 
skewness and flatness of velocity increments, performed as well and 
are consistent with previous experimental results. Furthermore,
the accumulation of energy in the small-scales associated with the
bottleneck seems to occur on a span of wavenumbers that is
independent of the Reynolds number, possibly ruling out an inertial
range explanation for it. Finally, intermittency exponents seem to
depart from standard models at high $R_{\lambda}$, leaving the
interpretation of intermittency an open problem.
\end{abstract}
\maketitle

\section{Introduction}
Turbulence prevails in the universe, and its multi-scale properties
affect the global dynamics of geophysical and astrophysical flows at
large scale, e.g. through a non-zero energy dissipation even at
very high Reynolds number $R_e$. Furthermore, small-scale strong
intermittent events, such as the emergence of tornadoes and hurricanes
in atmospheric flows, may be very disruptive to the global dynamics
and to the structure of turbulent flows. Typically energy is supplied
to the flows in the large scales, e.g., by a large scale instability.
The flow at these scales is inhomogeneous and anisotropic. In
the standard picture of turbulence, the energy cascades to smaller
scales due to the stretching of vortices by interactions with similar
size eddies. It is then believed that at sufficiently small scales
the statistics of the flow are independent of the exact forcing
mechanism, and as a result, its properties are universal. For this
reason, typical investigations of turbulence consider flows that are
forced in the large scales by a random statistically isotropic and
homogeneous body force \cite{Haugen06,Kaneda03}. However, how fast
(and for which measured quantities) is isotropy, homogeneity, and
universality obtained is still an open question.

The return to isotropy has been investigated thoroughly in the past,
by analysis of data from experiments and direct numerical simulations
(DNS) \cite{Sreenivasan97,Shen00,Pope,Kurien00,Biferale01a,Biferale02}.
However, lack of computational power limited the numerical investigations
of anisotropic forced flows to moderate Reynolds numbers, for which a
clear distinction of the inertial range from the bottleneck, and
from the dissipative range, cannot be made. Only recently the fast
increase of computational power permitted DNS to resolve
sufficiently small scales, such that a flow due to an inhomogeneous
and anisotropic forcing develops a clear inertial range with constant
energy flux. As a result, this kind of questions can be addressed
anew. To give an estimate of the size of the desired grid,
we mention that in recent simulations \cite{Mininni06} an
incipient inertial range was achieved for a resolution of $1024^3$
grid points, while for a $512^3$ run the range of scales between the
large scale forcing and the bottleneck was much less than an order
of magnitude. In all cases, the flow was resolved since
$k_{max}/k_\eta \gtrsim 1$, with $k_{max}$ the maximum wavenumber in the
simulation and $k_\eta$ the dissipation wavenumber built on the Kolmogorov
phenomenology.

Of particular interest in the study of turbulent flows is the issue
of universality. It is now known that two-dimensional turbulence
possesses classes of universality \cite{Bernard06}, and at least for
linear systems such as the advection of a passive tracer, there is
evidence of universality of the scaling exponents of the
fluctuations \cite{Biferale04}. However, recent numerical
simulations of three dimensional turbulence \cite{Mininni06} showed
that scaling exponents of two different flows (one non-helical, the
other fully helical) were measurably different at similar Reynolds
number. It is yet unclear whether this is an effect of anisotropies
in the flow, or of a finite Reynolds numbers. If this is a finite
Reynolds number effect, one then needs to ask how fast its
convergence to the universal value is obtained. If the convergence
rate is sufficiently slow then finite Reynolds effects should be
considered when studying turbulent flows that appear in nature, at
very large but finite Reynolds numbers. Thus, the question of the
universal properties of turbulent flows at high Reynolds numbers
remains somewhat open.

The recovery of isotropy, the differences observed in the scaling
exponents, and the slow emergence of scaling laws have been recently
considered in the context of the influence of the large scales on
the properties of turbulent fluctuations
\cite{Laval01,Alexakis05b,Mininni06}. The study of nonlocal interactions
between large and small scales has been carried in experiments and in
simulations
\cite{Domaradzki88,Domaradzki90,Kerr90,Yeung91,Okhitani92,Zhou93,Zhou93b,Brasseur94,Yeung95,Zhou96,Kishida99,Carlier01,Verma05,Alexakis05b,Mininni06,Poulain06}
at small and moderate Reynolds numbers. In simulations with $1024^3$
grid points \cite{Alexakis05b}, it was found that although most of the
flux is due to local interactions, non-local interactions with the large
scale flow are responsible for $\approx 20 \%$ of the total flux. It is
however unclear how the amplitude of these interactions scale with the
Reynolds number.

In this context, we solve numerically the equations for an incompressible
fluid with constant mass density. The Navier-Stokes equation reads
\begin{equation}
\partial_t {\bf u} + {\bf u}\cdot \nabla {\bf u} = - \nabla {\cal P}
    + \nu \nabla^2 {\bf u} +{\bf F} ,
\label{eq:momentum} \end{equation}
with $\nabla \cdot {\bf u} =0 $,
where ${\bf u}$ is the velocity field, ${\cal P}$ is the pressure divided
by the mass density, and $\nu$ is the kinematic viscosity. Here, ${\bf F}$
is an external force that drives the turbulence. The mode with the largest
wavevector in the Fourier transform of ${\bf F}$ is defined as $k_F$, with
the forcing scale given by $2 \pi / k_F$. We also define the viscous
dissipation wavenumber as $k_\eta=(\epsilon/\nu^3)^{1/4}$, where
$\epsilon$ is the energy injection rate (as a result, the Kolmogorov scale
is $\eta = 2\pi/k_\eta$).

The results in the following sections stem from the analysis of a series
of DNS of Eq. (\ref{eq:momentum}) using a parallel pseudospectral
code in a three dimensional box of size $2\pi$ with periodic boundary
conditions, up to a resolution of $2048^3$ grid points. The
equations are evolved in time using a second order Runge-Kutta method,
and the code uses the $2/3$-rule for dealiasing. As a result, the
maximum wavenumber is $k_{max} = N/3$ where $N$ is the number of grid
points in each direction.

With $L$ and $\lambda$ defined as
\begin{equation}
L = 2\pi \frac{\int{E(k) k^{-1} dk}}{\int{E(k) dk}}, \ \ \
\lambda = 2\pi \left(\frac{\int{E(k) dk}}{\int{E(k) k^2 dk}}\right)^{1/2},
\label{eq:integral} \end{equation}
the  integral scale  and Taylor scale respectively, the Reynolds
number is $R_e = UL/\nu$ and the Taylor based Reynolds number is
$R_\lambda = U\lambda/\nu$. Here, $U=\left< {\bf u}^2 \right>^{1/2}$
is the r.m.s. velocity and $E(k)$ the energy spectrum. The large scale
turnover time is $T=U/L$. Note that, with these definitions, $R_e$ and
$R_\lambda$ used in this paper are larger than the ones stemming
from the definitions used by the experimental community (see e.g.,
\cite{Frisch}) by a factor of $2 \pi (3/5)^{1/2}\approx 4.87$.

\begin{table}
\caption{\label{table:runs}Parameters used in the simulations. $N$ is
         the linear grid resolution, $\nu$ the kinematic viscosity, $R_e$
         the Reynolds number, and $R_\lambda$ the Taylor based Reynolds
         number.}
\begin{ruledtabular}
\begin{tabular}{ccccc}
Run & $N$  &     $\nu$         & $R_e$ & $R_\lambda$ \\
\hline
I   & 256  &$2\times 10^{-3}$  &  675  &     300     \\
II  & 512  &$1.5\times 10^{-3}$&  875  &     350     \\
III & 1024 &$3\times 10^{-4}$  & 3950  &     800     \\
IV  & 2048 &$1.2\times 10^{-4}$& 9970  &    1300
\end{tabular}
\end{ruledtabular}
\end{table}

Simulations were done with the same external forcing (see Table
\ref{table:runs} for the parameters of all the runs), with $U\approx 1$
in all steady states. The forcing ${\bf F}$ corresponds to a Taylor-Green
(TG) flow \cite{Taylor37}
{\setlength\arraycolsep{2pt}
\begin{eqnarray}
{\bf F} &=& f_0 \left[ \sin(k_F x) \cos(k_F y)
     \cos(k_F z) \hat{x} - \right. {} \nonumber \\
&& {} \left. - \cos(k_F x) \sin(k_F y)
     \cos(k_F z) \hat{y} \right] ,
\label{eq:TG}
\end{eqnarray}}
\noindent
 where $f_0$ is the forcing amplitude, and $k_F=2$. The
turbulent flow that results has no net helicity, although local
regions with strong positive and negative helicity develop.

\section{\label{sec:2048}The slow emergence of a Kolmogorov-like scaling}

\begin{figure}
\includegraphics[width=8cm]{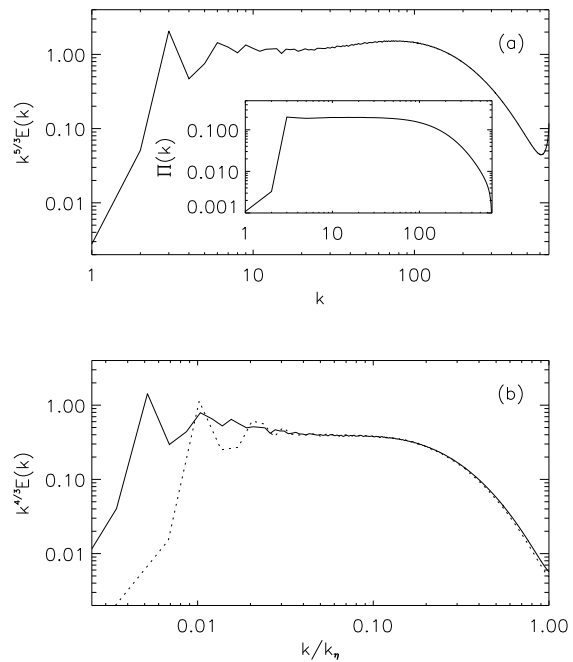}
\caption{(a) Energy spectrum in run IV compensated by $k^{-5/3}$. The
    inset shows the energy flux. (b) Energy spectrum in runs III (dotted)
    and IV (solid) compensated by $k^{-4/3}$. Wavenumbers are normalized
    by the dissipation wavenumber $k_\eta$.}
\label{fig:spectrum} \end{figure}

We first concentrate on the global dynamics of the $2048^3$ run (run IV).
Figure \ref{fig:spectrum}(a) shows the compensated energy spectrum in this
run, as well as the corresponding energy flux $\Pi(k)$, both taken in the
turbulent steady state after the initial transient. The energy flux is
constant in a wide range of scales, as expected in a Kolmogorov cascade,
but the compensated spectrum has a more complex structure in that same
range of scales. The salient features of this spectrum are well-known from
previous studies. Small scales before the dissipative range show the
so-called bottleneck effect with a slope shallower than $k^{-5/3}$. On
the other hand, larger scales have a spectrum with a slope slightly steeper
than $k^{-5/3}$, an effect that is even clearer in the simulation performed
at larger spatial resolution \cite{Kaneda03} on a grid of $4096^3$ points;
this small discrepancy with a Kolmogorov spectrum is attributed to
intermittency, i.e. to the spatial scarcity of strong events leading to
non-Gaussian wings in the probability distribution functions of velocity
gradients.

The bottleneck effect is not fully understood but clearly corresponds to
an accumulation of energy at the onset of the dissipation range. It
has been attributed to the quenching of local interactions close
to the dissipative scales \cite{Herring82,Falkovich94,Lohse95,Martinez97},
or to a cascade of helicity \cite{Kurien04} whose energy spectrum
would follow a $k^{-4/3}$ power law. The quenching of local interactions
in the bottleneck was measured directly in simulations in \cite{Mininni06},
and will be also shown here for run IV (see below, Figs.
\ref{fig:T3}-\ref{fig:scaling}). The $k^{-4/3}$ spectrum is also compatible
with the present data, as shown in Fig. \ref{fig:spectrum}(b) giving
the energy spectra in runs III and IV compensated by $k^{-4/3}$. However,
we observe that the width of the bottleneck appears to be independent of
the Reynolds number; this indicates that the origin of the bottleneck is
more likely a dissipative viscous effect than an inertial range effect.
If helicity plays a role in the formation of the bottleneck, it has to be
connected to the local generation of helicity at small scales due to the
viscous term in the Navier-Stokes equation. Purely helical structures are
exact solutions of the Navier-Stokes equation, and as a result an increase
of helicity in the small scales could quench local interactions and the
cascade rate (as assumed in Ref. \cite{Kurien04}).

\begin{figure}
\includegraphics[width=8cm]{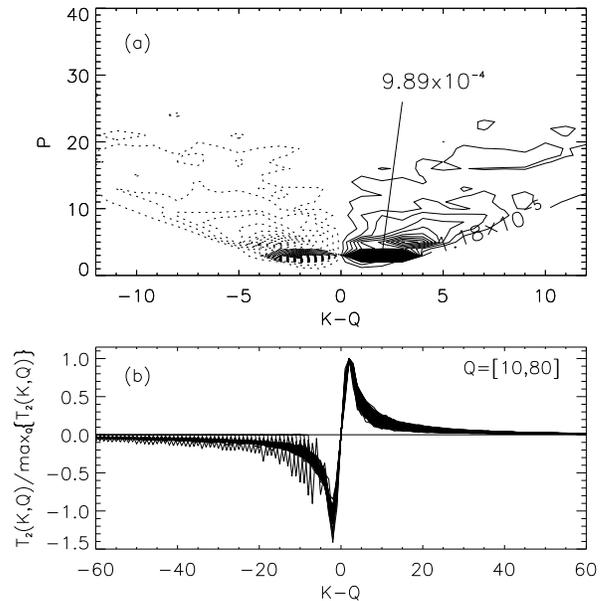}
\caption{(a) Amplitude of the triadic interactions $T_3(K,P,Q)$ for $Q=40$
    as a function of $K$ and $P$ in run IV. (b) Shell-to-shell energy
    transfer $T_3(K,Q)$ in the same run; several values of $Q$ are
    superimposed.}
\label{fig:T3} \end{figure}

The relative strength of local versus non-local interactions between Fourier
modes in the shell-to-shell transfer, and in the energy flux can be measured
in numerical simulations with the help of a variety of transfer functions
\cite{Kraichnan71,Lesieur,Verma04,Alexakis05,Alexakis05b}. Specifically, the
amplitude of the basic triadic interactions between the modes in shells $K$,
$P$ and $Q$ is defined as:
\begin{equation}
T_3(K,P,Q) = -\int {\bf u}_K \cdot ({\bf u}_P \cdot \nabla) {\bf u}_Q
    d{\bf x}^3 ,
\label{trans_eq3} \end{equation}
where the notation ${\bf u}_K$ denotes the velocity field filtered
to preserve only the modes in Fourier space with wavenumbers in the interval
$[K,K+1)$. Picking a wavenumber in the inertial range (here $Q=40$), we show
in Fig. \ref{fig:T3}(a) its amplitude as a function of $P$ and $K-Q$ for
run IV. Specific values of two levels are indicated as a reference (the
maximum, indicated by the arrow, corresponds to $P=k_F$). As a comparison,
in run III, $\max\{T_3(K,P,Q=40)\} \approx 1.4 \times 10^{-3}$ indicating that
a decrease of the relative amplitude of the non-local triadic interactions
with the large scale flow ($P=k_F$) occurs as the Reynolds number increases.
However, the non-local coupling of the modes with $P \approx k_F$ is still
dominant in run IV.

The relevance of these interactions in the transfer of energy
between scales can be quantified by studying the shell-to-shell transfer
and the net and partial fluxes. The energy transfer from the shell $Q$
to the shell $K$, integrating over the intermediate wavenumber, is defined as:
\begin{equation}
T_2(K,Q) = \sum_P T_3(K,P,Q) = -\int {\bf u}_K \cdot ( {\bf u} \cdot
    \nabla) {\bf u}_Q d{\bf x}^3 \ .
\label{trans_eq2} \end{equation}
It has the same qualitative behavior as in runs at lower
Reynolds number [see Fig. \ref{fig:T3}(b)]. The minimum of $T_2$ for
$K-Q \approx -k_F$ for all values of $Q$, and the maximum for
$K-Q \approx k_F$, both denote that the energy is transfered from the nearby
shell $K-k_f$ to the $Q$ shell, and transfered from this shell to the nearby
shell $K+k_f$. As a result, as we increase the Reynolds number, the
shell-to-shell energy transfer is still local but not self-similar,
mediated by strong non-local triadic interactions with the large scale
flow at $k_F$ \cite{Domaradzki90,Ohkitani92,Zhou93,Yeung95,Verma04,Alexakis05b,Mininni06,Poulain06}.

\begin{figure}
\includegraphics[width=8cm]{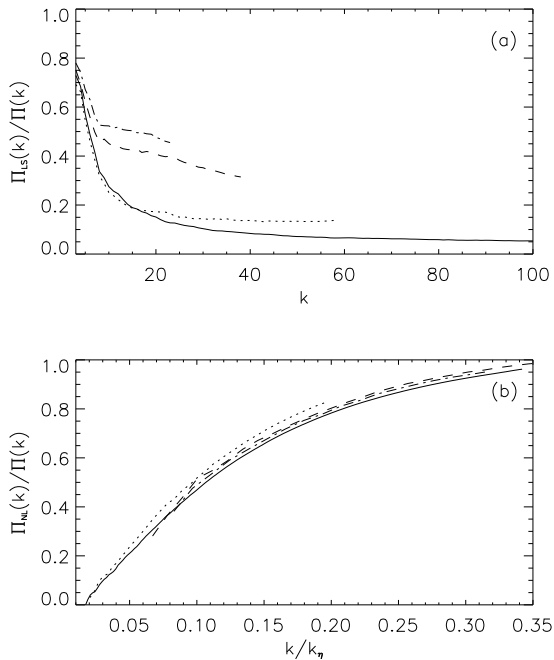}
\caption{(a) Ratio of large-scale to total energy flux
    $\Pi_{\textrm LS}(k)/\Pi(k)$ as a function of wavenumber for
    runs I (dash-dot), II (dash), III (dot), and IV (solid).
    (b) Ratio of nonlocal to total energy flux $\Pi_{\textrm NL}(k)/\Pi(k)$
    for the same runs; wavenumbers are in units of $k_\eta$. The fluxes are 
    defined in Eqs. (\ref{eq:flux}), (\ref{eq:fluxls}), and 
    (\ref{eq:fluxnl}).}
\label{fig:ratio} \end{figure}

It has been observed that although the individual non-local triadic
interactions are strong, as modes are summed to obtain the energy
flux, non-local effects become less relevant. To quantify further the net
contribution of the local and non-local effects to the energy flux,
we introduce the total flux
\begin{equation}
\Pi(k) = -\sum_{K=0}^k \sum_P \sum_Q T_3(K,P,Q) ,
\label{eq:flux}
\end{equation}
the energy flux due to the non-local interactions with {\it only} the
large scale flow
\begin{equation}
\Pi_{\textrm LS}(k) = -\sum_{K=0}^k \sum_{P=0}^{6} \sum_Q T_3(K,P,Q)
, \label{eq:fluxls}
\end{equation}
and the energy flux due to all the interactions outside the octave
around wavenumber $k$ (i.e., all non-local interactions)
\begin{equation}
\Pi_{\textrm NL}(k) = -\sum_{K=0}^k \sum_{P=0}^{k/2} \sum_Q T_3(K,P,Q) .
\label{eq:fluxnl}
\end{equation}

Figure \ref{fig:T3}(a) shows the ratio $\Pi_{\textrm LS}(k)/\Pi(k)$ as
a function of wavenumber for run IV. The same ratio for the lower
resolution runs in Table \ref{table:runs} are also shown here as a
reference. If the cascade is due to local interactions, this ratio
should decrease as smaller scales are reached. We observe however that,
at small scales, a plateau obtains within which this ratio remains
relatively constant. This is observed in runs III and IV, the two
runs at the highest Reynolds numbers. Note also that the plateau
lengthens as $R_{\lambda}$ increases: the length of the plateau
corresponds roughly to the length of the inertial range (including the
bottleneck) at those Reynolds numbers. Finally, the amplitude of the
plateau decreases as the Reynolds number is increased, indicating a
smaller contribution of the interactions with the large scale flow,
relative to the total flux. A detailed study of its dependence with
Reynolds number is discussed in the next section.

As previously mentioned, the ratio $\Pi_{\textrm LS}/\Pi$ does not
increase in the range of wavenumbers that spans the bottleneck. It is the
contribution of all non-local interactions (interactions with all the modes
outside the octave around a given wavenumber $k$) that becomes dominant in
this range. Figure \ref{fig:T3}(b) shows the ratio
$\Pi_{\textrm NL}(k)/\Pi(k)$ for the runs in Table \ref{table:runs} (note 
the wavenumbers are plotted in units of $k_\eta$). As scales closer to the 
dissipative range are considered, the contribution of all the non-local 
interactions increases, in agreement with the findings in Ref. 
\cite{Herring82}. Moreover, the amplitude of $\Pi_{\textrm NL}(k)/\Pi(k)$ 
when $k$ is in units of $k_\eta$ is roughly independent of the Reynolds, 
in agreement with a width of the bottleneck independent of $R_e$ and 
controlled by the growth of $\Pi_{\textrm NL}(k)$ as $k$ gets closer to 
the dissipation scale.

It is worth noting that even at the highest Reynolds number examined here,
there is still a significant contribution of nonlocal interactions
($\Pi_{\textrm LS}$ and $\Pi_{\textrm NL}$) to the total energy flux in
the inertial range. The comparison with runs at smaller resolution shows
a qualitative agreement and the persistence of the described non-local
effects. What the new computation at $R_{\lambda}\sim 1300$ allows,
though, is to determine the scaling of the relative importance of nonlocal
effects in Navier-Stokes turbulence when the Reynolds number is increased,
as we discuss next.

\section{Scaling laws in turbulent flows}

\begin{figure}
\includegraphics[width=8cm]{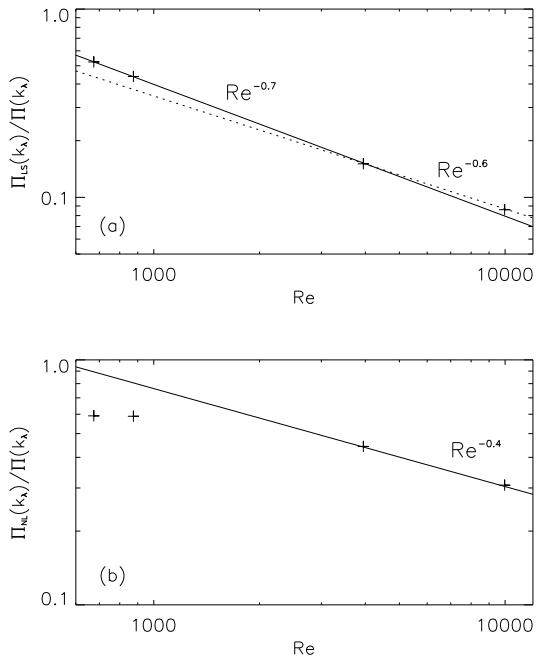}
\caption{Scaling of (a) the flux ratio $\Pi_{\textrm LS}/\Pi$ and (b) the
    non-local flux ratio $\Pi_{\textrm NL}/\Pi$ as a function of Reynolds
    number. Both ratios are evaluated at the Taylor scale, and several
    slopes are indicated as references.}
\label{fig:scaling} \end{figure}

Numerical simulations do not excel in the determination of scaling
laws in turbulent flows. The resolutions allowed by present day computers
barely allow for the existence of a well-defined inertial range. Indeed,
the observation of Fig. \ref{fig:spectrum} shows that, at this Taylor
Reynolds number, the Kolmogorov inertial range covers less than one order
of magnitude in scale (although, as noted before, the flux is constant in
a larger range of scales). This could be an indicator that solutions more
complex than simple power laws hold in the inertial range  \cite{Tsuji04}.
The pioneering computations of the Japanese group on the Earth Simulator
using random forcing has allowed, however, for some scaling laws to emerge,
although, as these authors observed, not all physical quantities of
interest converge to asymptotic values at the same rate
\cite{Ishihara05,Kaneda06}. We here
display such scaling laws for the particular flow studied, namely the 
Taylor-Green flow, relevant to several laboratory experiments. In 
particular, we are interested in the scaling of the relative amplitude 
of local and non-local interactions, as well as other quantities often 
studied in the context of turbulent flows, whose scaling will be used 
as a criteria to classify the runs \cite{Vanatta80,Ishihara05,Kaneda06}. 
It is worth mentioning in this context that, with four runs, we can only 
show the results are consistent (or at least, not inconsistent) with 
a particular scaling.

\begin{figure}
\includegraphics[width=8cm]{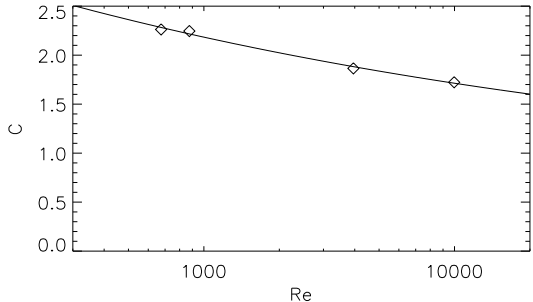}
\caption{Kolmogorov constant $C$ as a function of Reynolds number; the
 solid line gives the best fit $C = 4.60  R_e^{-0.16} + 0.64$.}
\label{fig:constant} \end{figure}

\begin{figure}
\includegraphics[width=8cm]{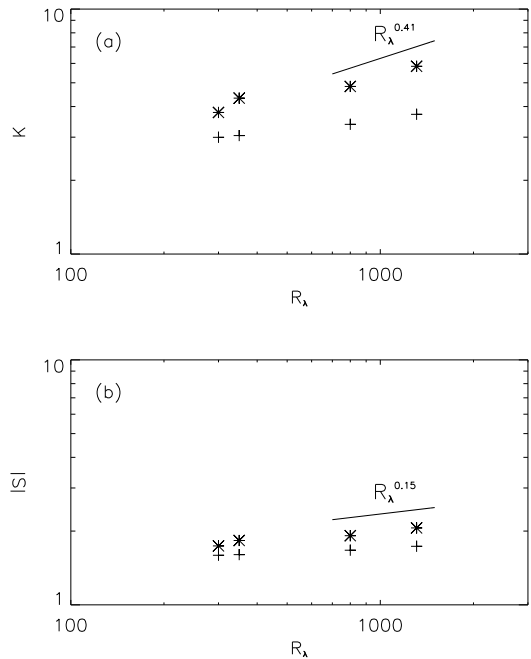}
\caption{(a) Kurtosis and (b) skewness of the velocity increments as a
    function of the Taylor based Reynolds number. Results are given for
    two different increments: the Taylor scale ($+$), and the
    dissipation scale ($*$). The slopes indicated as a reference are
    from experimental results.}
\label{fig:skewness} \end{figure}

Figure \ref{fig:scaling}(a) gives the scaling of the flux ratio
$\Pi_{\textrm LS}(k)/\Pi(k)$ with the Reynolds number. To this end,
we take the Taylor scale $\lambda$ as a reference scale in the
inertial range, and we evaluate $\Pi_{\textrm LS}(k)/\Pi(k)$ for
each run at the Taylor wavenumber $k_\lambda = 2 \pi/\lambda$. The
best fit to all the runs gives $\Pi_{\textrm LS}/\Pi \sim R_e^{-0.7}$,
although the dependence of the ratio $\Pi_{\textrm LS}/\Pi$ with $R_e$
seems to change slightly for Run IV. A best fit of the last two points
(runs III and IV) gives a dependence $\sim R_e^{-0.6}$ (as was
discussed in Sect. \ref{sec:2048}, these two runs show a developed
inertial range).

We also evaluate the ratio $\Pi_{\textrm NL}/\Pi$ at the Taylor
wavenumber; its dependence with the Reynolds number is shown in Fig.
\ref{fig:scaling}(b). Here, the ratio in runs III and IV is
compatible with a slower decay $\sim R_e^{-0.4}$. The anomalous
behavior of runs I and II in Fig. \ref{fig:scaling}(b) is due to the
fact that in these runs at lower resolution, the sum over $P$ from
the smallest wavenumber $k_{min}=1$ to $k_\lambda/2$ in Eq.
(\ref{eq:fluxnl}) defines bands that are too narrow in Fourier
space. In other words, it is linked to the lack of a well-defined
inertial range in the simulations at lower Reynolds numbers, and we
can only expect scaling to obtain in the limit of large $R_e$.

Figure \ref{fig:scaling} indicates that as the Reynolds number is
increased, the contribution of the non-local interactions with the
large scale flow to the total flux decreases (as well as the
contribution of all non-local interactions, albeit at a slower
rate). On dimensional grounds $\Pi_{\textrm LS}(k_\lambda) \sim
U_L u_\lambda^2/L$. Here, $U_L$ is a characteristic velocity at the
large scale $L$, and $u_\lambda$ is a characteristic velocity at the
Taylor scale (note that this relation does not take into account
that structures are in fact multiscale \cite{Alexakis05b}). On the
other hand, for $\Pi_{\textrm LS}/\Pi \ll 1$, we have
$\Pi(k_\lambda) \sim u_\lambda^3/\lambda$. As a result, we may
expect $\Pi_{\textrm LS}(k_{\lambda})/\Pi(k_\lambda) \sim
R_e^{-1/2}$. The condition $\Pi_{\textrm LS}/\Pi \ll 1$ is not
satisfied in the simulations at lower resolution, and it is unclear
whether the departure of Run IV in Fig. \ref{fig:scaling} represents
a convergence to a different scaling than $\sim R_e^{-0.7}$ at very
large Reynolds numbers. This will require further studies at higher
numerical resolutions, a feat reachable with petascale computing.

Other scaling laws can be observed in this series of runs; in
particular, it is worth comparing the scaling of quantities for
which data exist from laboratory experiments or from previous
simulations. Figure \ref{fig:constant} shows the Kolmogorov constant
$C_K$ as defined by the inertial range spectrum $E(k)=C_K
\epsilon^{2/3} k^{-5/3}$. As a reference, we computed a best fit of
the form $C_K=a R_e^b + c$, as suggested e.g. in \cite{Tsuji04}, and
obtained $a=4.60$, $b=-0.16$, and $c=0.64$. The value of $c$
(that represents the asymptotic value of the Kolmogorov
constant for infinite $R_e$) obtained from this fit is in good
agreement with experimental results and atmospheric observations
\cite{Mydlarski96,Tsuji04}, although the values of $a$ and $b$
differ. We also note that the measured value of the Kolmogorov
constant for the $2048^3$ runs is more than double the value of the
expected asymptotic limit $c$, indicating that we are still far away
from an asymptotic behavior for large $R_e$.

Figure \ref{fig:skewness} shows the skewness
\begin{equation}
S=\left< \delta u_L(r)^3 \right> / \left< \delta u_L(r)^2 \right>^{3/2},
\end{equation}
and kurtosis
\begin{equation}
K=\left< \delta u_L(r)^4 \right> / \left< \delta u_L(r)^2 \right>^{2},
\end{equation}
of the longitudinal velocity increment $u_L= {\bf u} \cdot {\bf r}/r$
\begin{equation}
\delta u_L(r) = u_L({\bf x+r}) - u_L({\bf x}),
\end{equation}
i.e. the component of the velocity
in the direction of the increment. The skewness and kurtosis were
evaluated at two scales, $r=\lambda$, the Taylor scale, and $r=\eta$,
the dissipation scale. In the latter case, only the results from runs III
and IV show a dependence with $R_\lambda$ which is consistent with
experimental results \cite{Vanatta80}. The behavior of these two runs
further confirms that high Reynolds numbers are needed to observe scaling of
turbulent quantities.

\section{Intermittency and structures}

The Taylor-Green flows computed here correspond to an experimental
configuration of two counter-rotating cylinders, studied in the
laboratory for fluid turbulence as well as in the context of the
generation of magnetic fields in liquid metals. These flows
present both inhomogeneities and anisotropies in the large scales,
a resolved inertial range followed by a bottleneck, and a dissipative
range. One may study the rate at which the symmetries of the Navier-Stokes
equations are recovered in the small scales, and whether the statistical
properties of the small scales are universal. In this section we address
the specific question of the properties of the small scales through the
evaluation of the anomalous exponents $\zeta_p$ of the longitudinal
structure functions $S_p$ of the velocity field, defined as:
\begin{equation}
S_p= \langle \delta u_L(r)^p \rangle \sim r^{\zeta_p},
\end{equation}
assuming homogeneity and isotropy. In order to obtain better scaling
laws, we use the Extended Self-Similarity hypothesis (ESS)
\cite{Benzi93,Benzi93b} in the particular context of plotting $S_p$
as a function of $S_3$.

\begin{figure}
\includegraphics[width=8cm]{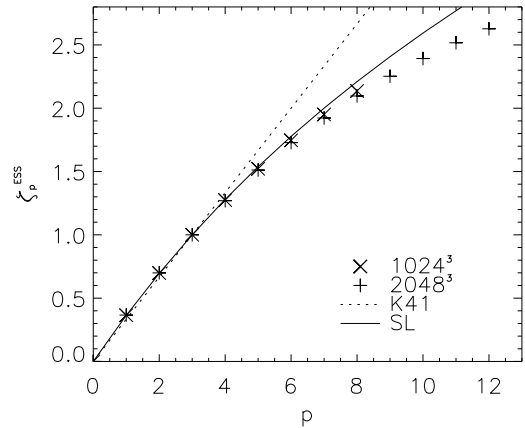}
\caption{Scaling exponents using the ESS hypothesis in the $1024^3$ and
    $2048^3$ runs. The scaling predicted by Kolmogorov and by the
    She-L\'ev\^eque model are also given as a reference.}
\label{fig:exponents} \end{figure}

Figure \ref{fig:exponents} shows the scaling exponents $\zeta_p$ in the
$1024^3$ and $2048^3$ runs, computed using the ESS hypothesis. Similar
results are obtained without ESS and doing the fit only in the inertial
range, defined as the range of scales where the so-called 4/5th law of
Kolmogorov 
is satisfied, namely $S_3(r) \sim r$. If we
define stronger intermittency as stronger departure from the Kolmogorov
scaling $\zeta_p=p/3$, we note that as we increase the Reynolds number,
the intermittency increases as well, albeit slowly. Furthermore, for
higher $R_{\lambda}$ (run IV), the departure from the She-L\'ev\^eque
model \cite{She94} increases (compared with run III), even for fixed
values of $p$. The differences between $\zeta_p$ for runs III and
IV, albeit small, are at least one order of magnitude larger than the
errors in the fit using ESS. As an example, in run III
$\zeta_6=1.746 \pm 0.003$ and $\zeta_8=2.136 \pm 0.007$, while in run IV
$\zeta_6=1.7284 \pm 0.0004$ and $\zeta_8=2.0968 \pm 0.0007$.

Here it is worth separating the discussion in two parts. On the one
hand, the increase of the departure from the She-L\'ev\^eque model
as the Reynolds number and spatial resolution are increased
indicates that the departure is not the result of lack of
statistics. This change in the exponents for simulations with the
same forcing at different Reynolds numbers shows that huge Reynolds
are required to obtain convergence of high order statistics. In fact, 
the larger the moment $p$ examined, the larger the relative difference 
between the $\zeta_p$ exponents measured in the two runs. 
On the other hand, it was shown in Ref. \cite{Mininni06} that
differences in the scaling exponents were measurable when
considering two different forcings at similar Reynolds numbers.
These differences could be due to anisotropies in the flow, and in
that case an SO(3) decomposition could be used to study whether the
scaling exponents of the isotropic component of the flow are
universal. However, if there is a significant return to isotropy in
the small scales, we then also expect the isotropic component to
dominate when the Reynolds number is large enough.

\begin{figure}
\vskip0.3truein \includegraphics[width=8.7cm]{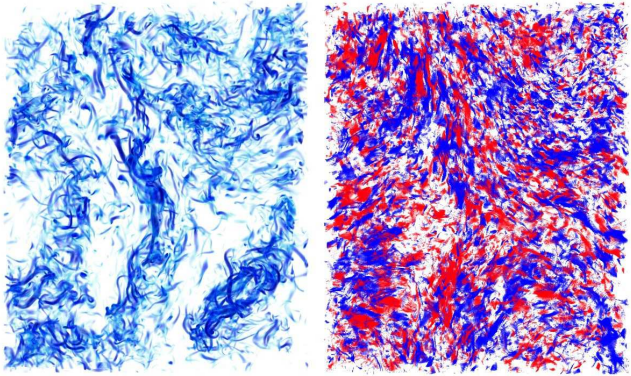}
\caption{(Color online) Left: rendering of vorticity intensity in a small
    region of run IV. Only regions with
    $|\boldsymbol{\omega}| \ge \max\{|\boldsymbol{\omega}|\}/6.5$ are shown
    ($\boldsymbol{\omega}=\nabla \times {\bf v}$). Note the clustering of
    filaments into larger vorticity structures. The bars on
    the bottom indicate respectively the integral, Taylor, and dissipation
    scales. Right: rendering of relative helicity in the same region (red
    is $-1$ and blue is 1). Only regions with absolute value larger than
    $0.92$ are shown.}
\label{fig:structure} \end{figure}

The intermittency of the flow is linked to the presence of strong
spatially separated structures in the form of vortex filaments. The high
$R_{\lambda}$ computation (run IV) displays the same large-scale
structure of bands as the run presented in \cite{Mininni06}. Conditional
statistics analysis as the ones performed in \cite{Mininni06} keep
showing a correlation between large scale shear and small scale gradients
and enhanced intermittency. It has been noted by several authors that
filaments tend to cluster into larger filamentary structures; this is
observed e.g. for supersonic turbulence \cite{Porter98} and in the
interstellar medium,
and it has been analyzed quantitatively in \cite{Moisy04}. When
individual structures are studied in real space, filament-like clusters
formed by smaller vortex filaments are observed here again (see Fig.
\ref{fig:structure}), something that was not seen in simulations of
the TG flow at lower resolution. This could be interpreted as a
manifestation of self-similarity, and a more quantitative analysis will
be presented elsewhere. In particular, it would be of interest to compute
the inter-cluster distance, and the intra-cluster inter-filament distance,
to see whether the space-filling factor of such flows diminish with
increasing Reynolds number. Note that the vortex cluster reaches a
global length comparable to the integral scale of the flow (indicated
in Fig. \ref{fig:structure}); as such, they may be a real-space
manifestation of the trace of non-local interactions between
small-scales (dominated by vortices) and large scales (dominated by the
forcing), giving a coherence length to the flow.

Figure \ref{fig:structure} also shows the density of relative
helicity ${\bf v} \cdot \boldsymbol{\omega} (|{\bf
v}||\boldsymbol{\omega}|)^{-1}$ ($\boldsymbol{\omega}=\nabla \times
{\bf v}$). Regions in blue and red correspond respectively to
regions of maximum alignment or anti-alignment between the two
fields (only regions with absolute relative helicity larger than
$0.92$ are shown). Note that regions with large relative helicity
correspond to small vortex tubes, but the filament-like clusters
have no coherent helicity. Regions with strong alignment fill a
substantial portion of the subvolume, even though the global
(relative) helicity of the flow is close to zero.

\section{Discussion and Conclusion}

The data presented in this paper has allowed for a refined analysis of
the behavior and structure of turbulent flows as the Reynolds number is
increased. We have in particular showed that: (i)  the bottleneck appears
to have a constant width for the two higher $R_e$ runs;
hence, it is probably linked to the dissipation range, and to the
depletion of nonlinearities as we approach this range; (ii) the
scaling with $R_e$ of the non-local energy fluxes, which indicates a
weakening of non-local interactions as $R_e$ increases. These first
two results taken together point out to the fact that the bottleneck
may not disappear in the limit of very high Reynolds number, since
it has been argued that its existence is linked to the relative
scarcity of non-local interactions in Navier-Stokes turbulence, by
opposition to, e.g., the magnetohydrodynamic (MHD) case. Indeed,
when coupling the velocity to a magnetic field in the
MHD limit, it was shown that the transfer of
energy itself was non-local, and that the bottleneck was absent in
numerical simulations of such flows; this can be understood in the
following manner: as one approaches the dissipation range, few
triadic interactions are available but in a flow for which the
nonlinear transfer is nonlocal, the energy near the dissipative
range can still be transfered efficiently to smaller scales since
small-scale fluctuations are transfered by the large scales
\cite{Alexakis05}. Finally, the departure of the anomalous exponents
of velocity structure functions from standard models of
intermittency such as the She-L\'ev\^eque model seems to increase as
the Reynolds number is increased.

As noted before in \cite{Kaneda06}, convergence to the asymptotic
turbulence regime appears to be very slow: even though the nonlocal
interactions do diminish with Reynolds number, they are still measurable
at these resolutions. In run IV on a $2048^3$ grid at
$R_{\lambda}\sim 1300$, of the order of $10\%$ of the energy flux is
due to non-local interactions with the large scale flow, and the
dependence of the energy flux ratio $\Pi_{\textrm LS}/\Pi$ with $R_e$ for
very large $R_e$ is still unclear.
This not only raises the question of the determination of higher
order quantities at moderate Reynolds numbers in simulations and
experiments, but it also opens the door for a non-universal behavior
of turbulent flows which may have to be studied in more detail than
was previously hoped for.

\begin{acknowledgments}
Computer time was provided by NCAR and by the National Science
Foundation Terascale Computing System at the Pittsburgh Supercomputing
Center. PDM and AP acknowledge invaluable support from Raghu Reddy at
PSC. PDM acknowledges discussions with D.O. G\'omez. PDM is a member 
of the Carrera del Investigador Cient\'{\i}fico of CONICET. AA 
acknowledges support from Observatoire de la C\^ote d'Azur and Rotary 
Club's district 1730. The NSF grant CMG-0327888 at NCAR supported this 
work in part. Three-dimensional visualizations of the flows were done 
using VAPOR, a software for interactive visualization and analysis of 
terascale datasets \cite{Clyne07}. 
\end{acknowledgments}

\bibliography{ms}

\end{document}